\documentclass[lettersize,journal]{IEEEtran}
\usepackage{amsmath,amsfonts}
\usepackage{algorithmic}
\usepackage{algorithm}
\usepackage{array}
\usepackage[caption=false,font=normalsize,labelfont=sf,textfont=sf]{subfig}
\usepackage{textcomp}
\usepackage{stfloats}
\usepackage{url}
\usepackage{verbatim}
\usepackage{graphicx}
\usepackage{xcolor}
\usepackage{cite}
\usepackage{enumitem}
\setlist[itemize]{leftmargin=2em}
\setenumerate[1]{itemsep=0pt,partopsep=0pt,parsep=\parskip,topsep=5pt}
\setitemize[1]{itemsep=0pt,partopsep=0pt,parsep=\parskip,topsep=5pt}
\setdescription{itemsep=0pt,partopsep=0pt,parsep=\parskip,topsep=5pt}
\begin{document}

\title{Multi-Player Immersive Communications and Interactions in Metaverse: Challenges, Architecture, and Future Directions}

\label{author}
\author{
	Yakun~Huang,
	Xiuquan~Qiao,
	Haowen~Wang,
	Xiang~Su,~\IEEEmembership{Member,~IEEE,}\\
	Schahram~Dustdar,~\IEEEmembership{Fellow,~IEEE,}
	Ping~Zhang,~\IEEEmembership{Fellow,~IEEE}
	\IEEEcompsocitemizethanks{
		\IEEEcompsocthanksitem  Y. Huang, X. Qiao, H. Wang and P. Zhang are with the State Key Laboratory of Networking and Switching Technology, Beijing University of Posts and Telecommunications, Beijing 100876, China. E-mail:\{ykhuang, qiaoxq, hw.wang, pzhang\}@bupt.edu.cn.
		\IEEEcompsocthanksitem X. Su is with the Department of Computer Science, Norwegian University of Science and Technology, 2815 Gj{\o}vik, Norway and University of Oulu, 90570, Oulu, Finland. Email:xiang.su@ntnu.no.
		\IEEEcompsocthanksitem S. Dustdar is with the Distributed Systems Group, Technische Universität Wien, 1040 Vienna, Austria. E-mail:dustdar@dsg.tuwien.ac.at.
	}
\vspace{-0.8cm}
}
	

\markboth{Journal of \LaTeX\ Class Files,~Vol.~14, No.~8, October~2022}%
{Shell \MakeLowercase{\textit{et al.}}: A Sample Article Using IEEEtran.cls for IEEE Journals}


\maketitle

\begin{abstract}
The metaverse has awakened users' expectations of an immersive interaction that fuses the virtual digital world and the physical world across space and time. 
However, the metaverse is still in its infancy, typically expanding multi-player applications (e.g., multi-player games) to implement a prototype with the help of 5G/Beyond 5G, Artificial Intelligence, digital twin, and other enabling technologies.
This article reviews the characteristics, key enabling technologies, and driving applications of the state-of-the-art metaverse.
We focus on the immersive interactions perspective of the metaverse from the tasks, inputs, and feedback across the users, digital world, and physical world and reveal the key challenges.
Afterwards, we present a multi-player interaction prototype platform based on a cloud-edge-device collaborative framework.
Also, we evaluate it with centralized and device-to-device~(D2D) approaches to verify the efficiency and flexibility of interactions.
Finally, we point out future research approaches and discuss potential solutions to enable more stable and higher quality multi-player interactions for metaverse services.
\end{abstract}

\section{Introduction}

\IEEEPARstart{M}{etaverse} is an emerging paradigm for the next generation of Internet, where users live in the virtual digital world as if they were living in the physical world~\cite{tang2022roadmap, wang2022survey}.
Metaverse envisions an immersive perception, computation, and interaction platform that supports cross-temporal sharing and persistent online interaction leveraging enabling technologies, such as Extended Reality~(XR), Internet of Things~(IoT), Artificial Intelligence~(AI), and digital twin.
The metaverse connects users through digital worlds, establishes diverse metaverse zones~(i.e., a virtual interaction space focused on a specific topic) to serve different groups, and develops meaningful new social relationships.
However, the metaverse is still in its infancy, and existing research focuses on applications extending traditional interaction technologies, such as Virtual Reality~(VR), Augmented Reality~(AR), and Mixed Reality~(MR).
Also, they are gradually exploring multiple fields, such as multi-player games, business, education, and entertainment.
For instance, influenced by COVID-19, online virtual offices and seminars have been popular, facilitating users' communication and social interactions.

We investigate reviews, surveys, and magazine papers related to the metaverse from various perspectives.
Tang~\textit{et al.}~\cite{tang2022roadmap} describe the metaverse roadmap in terms of 6G communication and networking, revealing the stringent requirements and challenges of 6G implementation of the metaverse.
Zhao~\textit{et al.}~\cite{zhao2022metaverse} construct the basic metaverse framework from a visual interaction, including the visual construction of the digital world, visualization, and user interactions.
In addition, the literature~\cite{wang2022survey} introduces the metaverse from a security and privacy perspective and presents the key challenges of security and privacy threats.
Zhao~\textit{et al.}~\cite{yang2022fusing} investigate the integration of blockchain and AI with the metaverse for protecting different metaverse zones.
Most efforts explore integrating existing technologies with the metaverse, such as 6G, AI, and blockchain.
However, there is a lack of in-depth investigation on the challenges of multi-player interaction from a ``user-centered" perspective.

The metaverse exhibits complex, multimodal interaction services between users, the digital world, and the physical world. 
In particular, real-time data flow and multi-player interaction across different metaverse zones are the key challenges guaranteeing immersive experiences.
First, intelligent devices support the metaverse's powerful perception and interaction capabilities but require more intensive rendering computations.
For example, large-scale reconstruction and rendering, synchronization, and real-time interaction pose difficulties for multi-player interactions.
Second, heterogeneous metaverse services require users to synchronize and switch services smoothly between digital and physical worlds in different metaverse zones.
Last, flexible configuration and optimization of multidimensional network resources in a large and complex scenario is also significant in multi-player interaction.

In this article, we start with reviewing existing surveys on metaverse and then introduce the concept, features, key enabling technologies, applications, and challenges of multi-player interaction services.
Afterwards, we present an implementation of a cloud-edge-device collaborative architecture for efficient and flexible interactions, conduct experiments to validate the effectiveness, and pave the way for realizing the metaverse.
Our contributions are threefold: 1) We review the state of the art of the metaverse and reveal the challenges in multi-player interactions;
2) We propose and implement a prototype for multi-player interactions in metaverse;
3) We conduct extensive experiments for evaluation and pointed out future research directions.

\vspace{-0.2cm}
\section{Metaverse: Characteristics, Enabling Technologies, Applications, and Challenges}

\subsection{Concepts and Characteristics}
The metaverse allows users to control the digital elements and interact with other user-controlled elements in a virtual shared space, gaining interaction experiences comparable to those in the real world.
For instance, leveraging XR interaction devices, users control their digital avatars for daily communication, collaboration, sharing, and gaming in the metaverse zone.
Meanwhile, users can quickly build and travel through different metaverse zones and create digital assets.
Thus, the metaverse builds a deeply integrated and interactive immersive spatial platform around the physical world, the digital world, and humans, exhibiting the following key characteristics:
\begin{figure*}[htbp]
	\vspace{-0.3cm}
	\centering
	\includegraphics[width=0.85\textwidth]{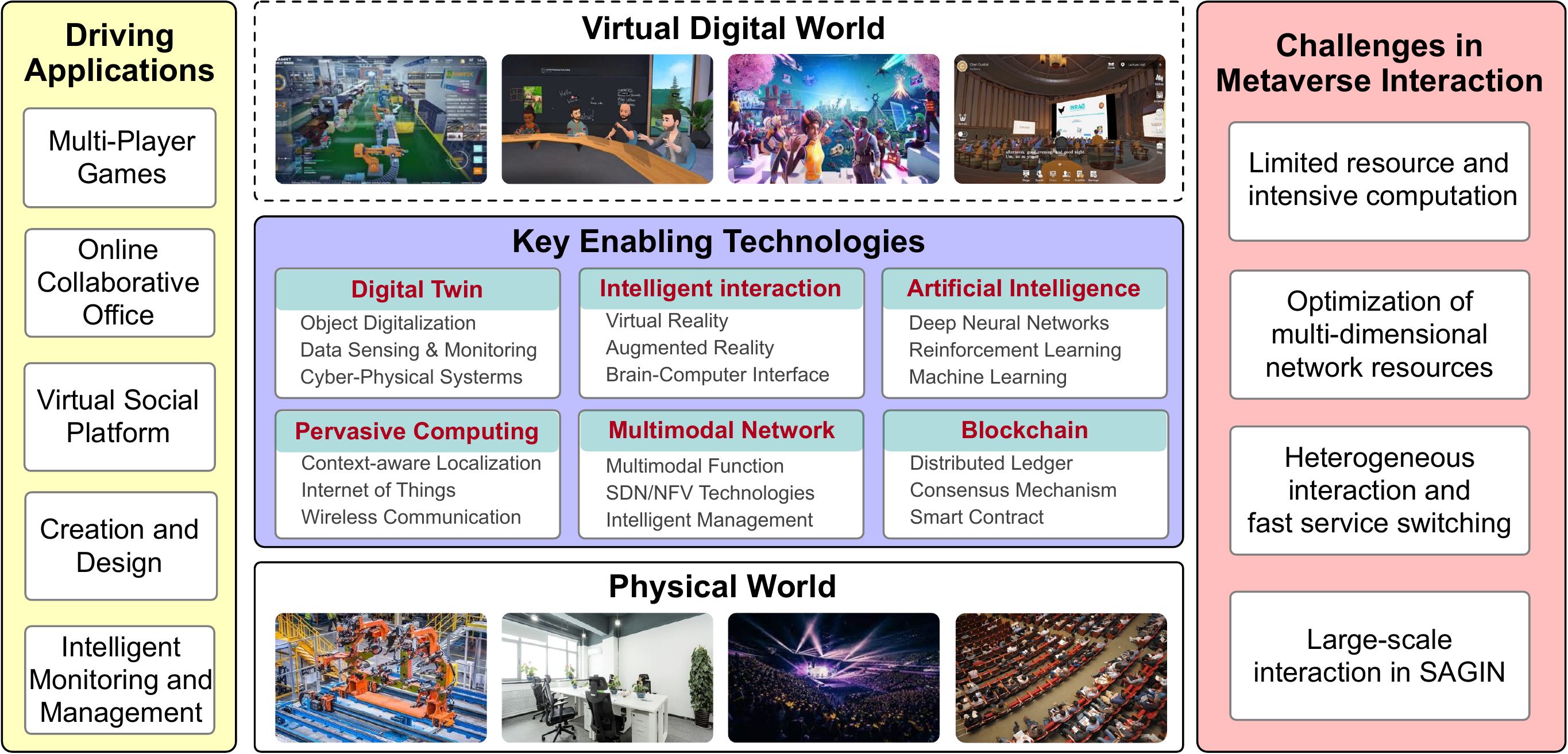}
	\caption{Driving applications, key enabling technologies, and challenges in Metaverse.}
	\label{Fig_1}
	\vspace{-0.3cm}
\end{figure*}

\paragraph{\textbf{Deep immersion}}
To achieve the ultimate integration with the physical world and allow users to be visually immersed, the virtual digital world in the metaverse requires realistic enough, such as digital avatar reconstruction.
Besides, the human digital body is enhanced by multimodal perception capabilities, including vision, hearing, touch, taste, brain signal, and myoelectricity~\cite{wang2022survey}.
Hence, users can obtain the immersion from the physiological perspective and synchronize, transmit and interact with these perceptual signals between different digital avatars.

\paragraph{\textbf{Spanning space-time}}
Existing contact in the physical world is limited by spatial distance and cannot turn back time, which limits real-time human interactions across space and time. 
In contrast, the metaverse is a hyperspace environment parallel to the physical world, which can break the time and space limitations~\cite{dionisio20133d}. 
For example, users can switch and enter the digital worlds in different spatio-temporal dimensions to interact with different users. 
Users can even quickly switch between life scenes, locations, and timelines, which are currently unattainable in the physical world.

\paragraph{\textbf{High interactivity}}
Users can travel and interact in different metaverse zones using the digital avatar.
They can also synchronize, interact, and give feedback on properties and states of the environment in the digital and physical world.
The digital assets created by the user can also interact with each other. 
Besides, different digital and physical worlds have various interaction tasks and themes in metaverse zones, and such tasks and states of the metaverse elements are ever-changing. 
Thus synchronizing interaction and giving feedback is an important feature of metaverse interaction.
 
\paragraph{\textbf{Flexibility and compatibility}}
The metaverse covers a variety of complex physical environments; for example, a user accesses the metaverse while traveling or in a factory.
In particular, heterogeneous interaction devices may differ in device interfaces, data types, communications, and application services, which therefore requires the metaverse to cope with different environments flexibly.
Also, interaction terminals require to be compatible with pervasive devices.

\paragraph{\textbf{Freedom of creation}}
The metaverse introduces stronger characteristics of creative space and freedom to users, providing a safe and independent user space. 
They can create digital assets in the digital world and generate economic value by circulating in different metaverse zones.
At the same time, the metaverse can continuously stimulate users' enthusiasm for digital content creation and open innovation.

\vspace{-0.2cm}
\subsection{Key Enabling Technologies}
The metaverse embraces six crucial enabling technologies, as shown Figure.~1, including:
\paragraph{\textbf{Digital Twin}}
The metaverse is built on a digital twin world in parallel with the physical world. 
Digital twin technology provides digital clones of various elements and multi-dimensional states in the physical world, thus reconstructing and rendering a high-fidelity digital world.
It can further autonomously predict and optimize in the digital world and then return the decision or warning results to the users and the physical world.
Therefore, with digital twin technology, we can use the digital world to predict, regulate, trace, and other functions, thus reducing the possible accidents and risks in the physical world.

\paragraph{\textbf{Intelligent Interactions}}
Intelligent XR interactive terminals~\cite{qiao2019web} are expected to become the dominant interaction terminals of the metaverse.
Intelligent interactions require the provision of an immersive viewing experience, such as VR, gesture and posture interaction like AR, and other emerging interactions.
Besides, Human-Computer Interaction~(HCI) (especially Brain-Computer Interface~(BCI)), myoelectric interaction, and other new intelligent interactions are promising interaction technologies.
For example, the user wears an intelligent terminal to collect data on the user's physiological characteristics and provide intelligent health monitoring.
BCI interfaces support users using brain signals to interact with the physical world, a significant scientific breakthrough that breaks the interaction limitations of using cell phones and laptops. 

\paragraph{\textbf{Artificial Intelligence}}
AI technology drives all aspects of the metaverse and learns from massive multimodal data streams, providing better resource allocation, service scheduling, and network provisioning.
AI algorithms can be developed to adjust computing and communication resources and customize the digital avatar individually.
In particular, AI technology is indispensable in digital twin reconstruction and rendering to obtain more realistic and efficient real-time driving and rendering.

\paragraph{\textbf{Perversive Computing}}
The computing space accommodated by the metaverse includes space, sky, and sea, which also implies broader pervasive computing.
For example, users achieve immersive interactive experiences based on lightweight wearable devices rather than high-performance computers.
With the collaborative capabilities of pervasive computing, offloading tasks flexibly and dynamically to pervasive devices and computing centers improve the efficiency of interactions.

\paragraph{\textbf{Multimodal Networks}}
The metaverse requires higher network bandwidth, a denser number of connections, and more reliable quality of service. 
6G, software-defined network~(SDN), SAGIN, and endogenous AI network service architecture are the key technologies that support the real-time transmission of massive data between digital and physical worlds~\cite{chen2021edge}.
In particular, SAGIN extends the network communication range and customizes the multimodal network services to the large-scale metaverse.
As a result, metaverse provides reliable and real-time sensing and transmission in different granularity through multimodal network services, consequently establishing a realistic and complete digital twin and realizing real-time interaction and feedback.

\paragraph{\textbf{Blockchain}}
The metaverse contains many private data in the physical world.
Therefore, preserving data privacy and security in the meta-universe is crucial for providing immersive services.
A decentralized architecture of interactions can avoid the risks of the traditional centralization method and keep the data persistent in the metaverse.
This means decentralized service provided by blockchain technology is an important foundation of the metaverse.
This distributed ledger constructs data into hash chain blocks for storage with important features such as decentralization, invariance, transparency, and auditability~\cite{wang2021blockchain}.
Therefore, blockchain technology can enable secure, transparent, and sophisticated services in the metaverse, such as virtual financial transactions.

\vspace{-0.2cm}
\subsection{Driving Applications}
In this section, we introduce some representative killer applications in the metaverse.
\paragraph{\textbf{Multi-player games}}
The concept prototype of the metaverse is based on multi-player interactive games, enriching interaction experiences with advanced technologies such as digital twins, AI, and pervasive computing.
For instance, Roblox is a representative multi-player interaction game platform where users can design and create items according to their preferences, such as skins, clothes, etc.
Fortnite is another online video game in which players can create worlds and battle arenas with their idea.

\paragraph{\textbf{Virtual social platform}}
Influenced by COVID-19, the virtual social platform has revolutionized how people socialize, enabling a range of immersive multi-player social applications, including virtual offices, virtual shopping, virtual concerts, and immersive cultural tours.
For example, Lil Nas X hosted a virtual concert on Roblox in 2020 that engaged over 30 million fans.
UC Berkeley virtually celebrated its graduation in Minecraft in 2020 by digitally replicating the campus landscape.

\paragraph{\textbf{Online collaborative office}}
Metaverse provides immersive collaborative experiences for multi-player online work across space limitations and enhances office efficiency.
Users can access the virtual offices or meeting rooms to discuss, learn or work through various terminals in the form of 2D video, digital avatars, or holographic videos.
Horizon Workrooms is Meta's office collaboration software (running in the Oculus Quest 2 headset) that allows users in any physical location to work and meet together in the same virtual world.
NetEase released an immersive system that holds academic meetings in virtual scenarios, supports setting up different styles, and can replicate meeting scenarios with high precision.

\paragraph{\textbf{\textbf{Creation and design}}}
The metaverse provides creators with unlimited possibilities for creation and imagination. 
Users can create and design a dazzling array of digital assets in the digital world without being limited by space and time. 
Also, these digital assets are supported to be exchanged among users and flow seamlessly in different metaverse zones.
More importantly, the metaverse can provide opportunities for experimentation in high-risk industrial applications, such as automotive and industrial product design, supporting pre-designing in a digital world before going into production.

\paragraph{\textbf{Intelligent monitoring and management}}
The metaverse has a ``smart brain" that links the digital and physical worlds by perceiving massive data, monitoring, and predicting future emergencies and dangers.
It greatly reduces the potential risks of users engaging in production activities and daily life in the physical world, thus better monitoring and managing all kinds of production.
For example, we can optimize the productions by analyzing the digital world's production indicators and historical records.
Another application is supply chain management, where we can rely on the digital world's data to analyze key performance, such as the actual efficiency of each process, and is particularly useful for optimizing production and scheduling processes in the physical world.

\vspace{-0.2cm}
\subsection{Challenges}
The metaverse provides users free and flexible access to immersive multi-player interaction services and satisfies real-time data flow, command synchronization, reconstruction, and rendering of digital worlds among various metaverse zones.
Therefore, realizing the above-mentioned exciting visions and application services poses new challenges to the metaverse, as manifested in the following aspects.

\paragraph{\textbf{Enabling massive rendering and interaction on resource-constrained devices}}
The metaverse access and immersive experience services leverage pervasive devices. 
However, pursuing lighter and more comfortable devices limits computing capability and battery life. 
Visual rendering and intelligent AI interaction in the metaverse requires intensive computing resources. 
Therefore, the first and foremost solution is to meet the demand of interaction latency to obtain a real-time response by offloading intensive computing to the edge or the cloud computers.

\paragraph{\textbf{Optimizing multi-dimensional network resources}}
The metaverse involves massive data sensing, transmission, computation, and rendering, which pose challenges to existing network services.
It is difficult to satisfy all requirements when providing personalized interaction using heterogeneous devices.
Especially when users dynamically shuttle in different metaverse zones, it is challenging to ensure the quality of interactions with different users by configuring, scheduling, and optimizing multi-dimensional network resources.

\begin{figure*}[!htbp]
    \vspace{-0.2cm}
	\centering
	\includegraphics[width=0.88\textwidth]{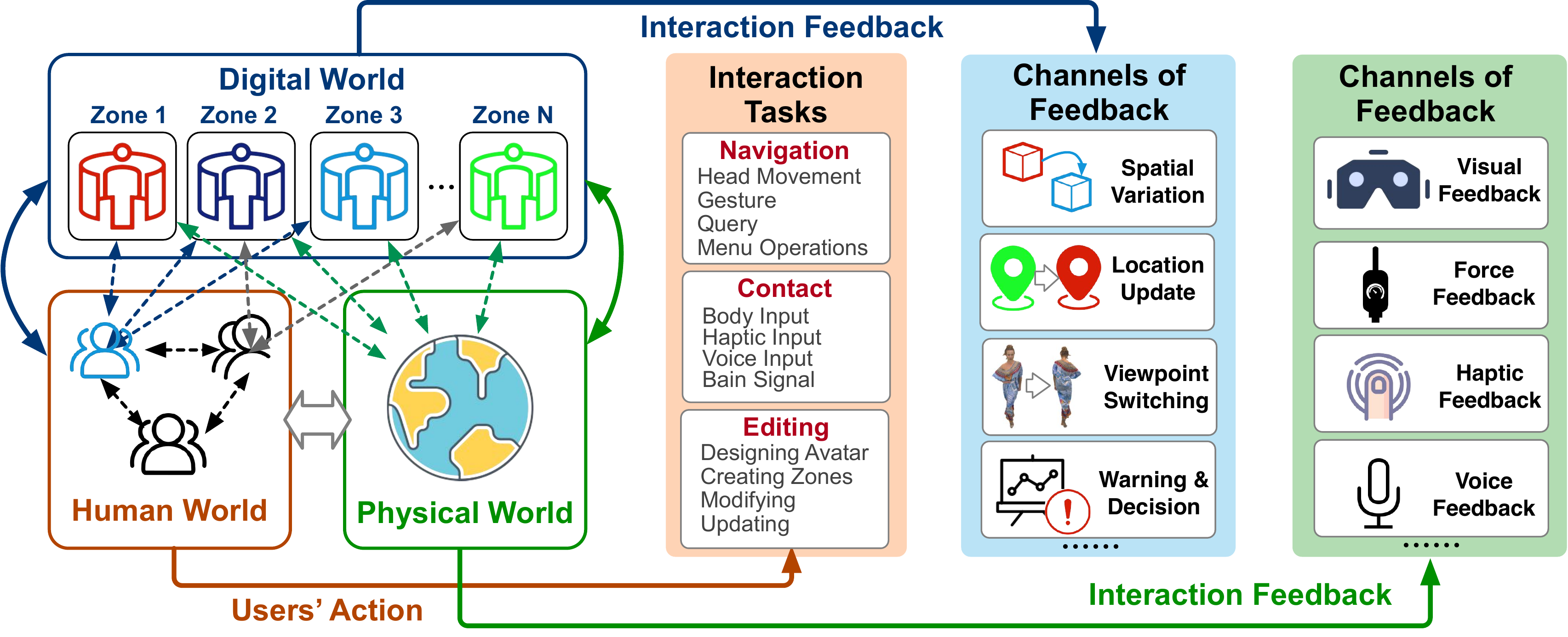}
	\caption{Interactions types and feedback channels in the metaverse.}
	\label{Fig_2}
	\vspace{-0.3cm}
\end{figure*}

\paragraph{\textbf{Heterogeneous interaction and fast service switching}}
Traveling through different metaverse zones with different digital avatars is the key to attracting user interest. 
Therefore, it is one of the supporting foundations to quickly complete service switching in heterogeneous metaverse zones and meet interaction requirements. 
Traditional service switching can be done quickly using virtual container techniques.
However, it is difficult to persist and quickly switch the interaction state of multi-player in the digital and physical world.
Also, maintaining multi-dimensional resources and states to satisfy different metaverse zones is challenging.

\paragraph{\textbf{Large-scale interaction in the sky-area integrated network~(SAGIN)}}
The metaverse can expand the physical distance of multi-player interactions, including air, sky, and sea. 
When involving a larger range of multi-player interactions, the fundamental technique guarantees the real-time transmission and synchronization of massive multi-dimensional sensory data and various metaverse zones.

\section{\textbf{Multi-player interactions in Metaverse}}
In this section, we focus on the interaction tasks in the multi-player metaverse.
Figure.~2 present three types of interactions and outlines the main perception and interaction feedback channels.

For a given interaction task, the user can get seamless interactions in different metaverse zones using physiological signal inputs and terminals for feedback output.
The general interaction tasks can be decomposed into three processes: navigation, contact, and editing~\cite{zhao2022metaverse}.
Interaction navigation refers to the user's current view operation, including geographic and non-geographic navigation.
Geographic navigation means the user is walking in the physical world or using an interaction terminal to move in the digital world.
Non-geographic navigation completes the navigation process by query, task, or other specified actions~\cite{jankowski2013survey}.
Contact interaction refers to the ability to touch and feel the physical world using various sensors, including direct and indirect approaches.
For example, in the direct contact approach, users can perceive the physical world through haptic and force feedback sensors and get a highly similar experience to the physical world. 
Indirect contact means users can control virtual physics in the digital world.
Editing interaction is modifying the editable element in the digital world, including dynamics and objects.
For example, users personalize their avatar or set the character of participating in different metaverse zones.

Next, we introduce the perception channels users interact with others in the metaverse, such as vision, touch, taste, and brain signals.
Electroencephalography (EEG), electromyography (EMG) biopotentials, and many other signals are gradually being recognized as potential future interactive control inputs in the metaverse, capable of liberating the user's interaction.
Then, users and the environment can obtain feedback from the visual screen or sensors of the intelligent interaction devices.
The feedback channels for interaction also include a variety of modalities, such as visual, auditory, and tactile.
For example, users can get an instant interactive experience through visual and audio feedback and communicate more realistically with virtual images in the digital world.
Haptic feedback can be obtained from various force feedback sensors, such as wearing smart XR gloves capable of simulating information about the user's shape and force, such as stiffness, friction, and sliding~\cite{bouzbib2021can}.

In addition, a more important type in the metaverse is collaborative interaction, consisting of communication, synchronization, and collaborative editing.
As shown in Figure.~2, when users are in the same metaverse space, they need to communicate and synchronize the information of different users in real time.
Also, collaborative editing is required when users interact across different metaverse zones.

\vspace{-0.2cm}
\section{Prototype and Evaluation}
\subsection{Prototype}
We present a multi-player interaction prototype to explore the provision of an efficient and flexible interaction service to meet a variety of potential future applications in the metaverse.
In the right part of Figure.~3, the proposed prototype is based on the cloud-edge-device collaboration architecture, which enables self-organized communications and interactions to satisfy different metaverse services for heterogeneous pervasive interaction devices.

\begin{figure*}[!htbp]
    \center
	\includegraphics[width=0.9\textwidth]{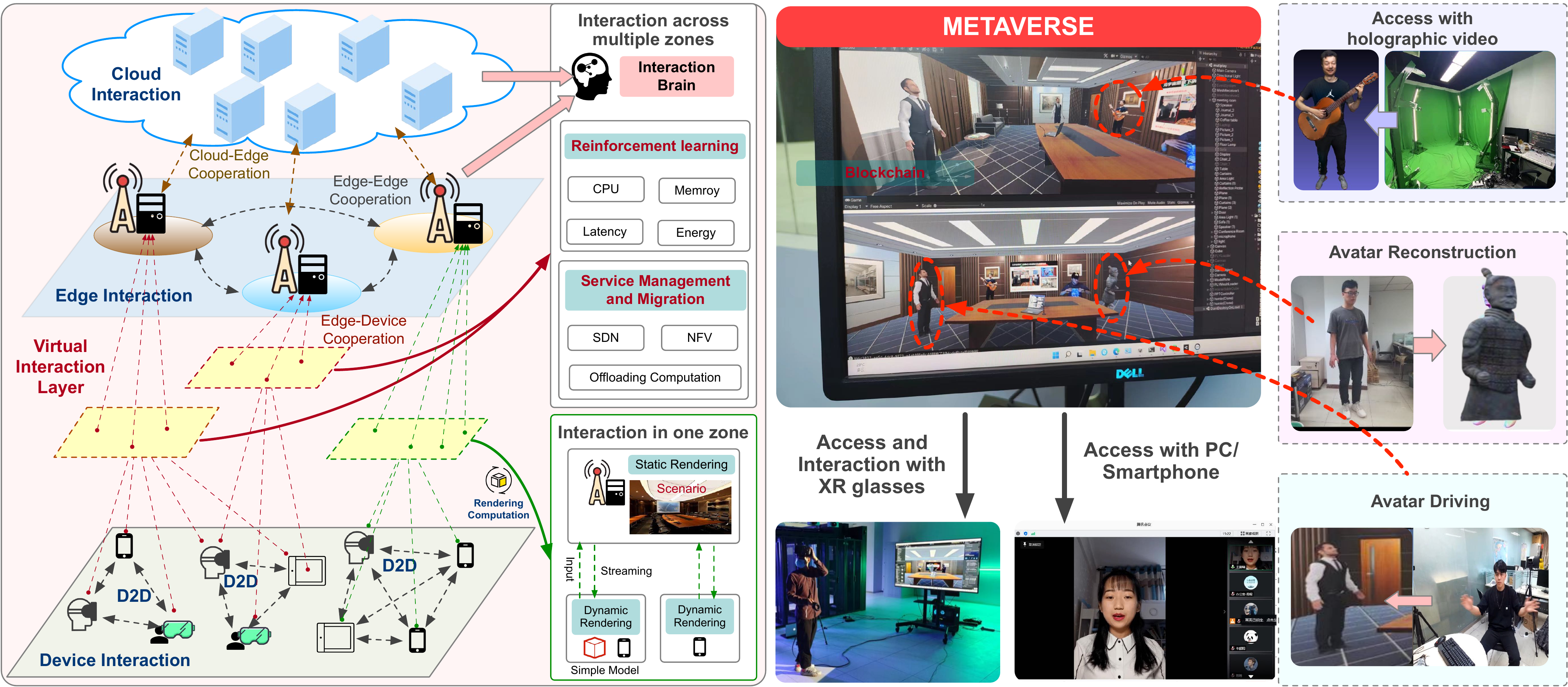}
	\caption{Prototype system of multi-player immersive interaction.}
	\label{Fig_3}
	\vspace{-0.4cm}
\end{figure*}

\textbf{Intelligent interaction layer.}
The terminal interaction layer is a direct entry point for access and interaction in the metaverse, e.g., a user can open an XR device and quickly choose to join the interested metaverse zone.
Figure.~3 exemplifies virtual interaction links in the interaction layer, including direct communication between terminals within a limited range in the physical world and interactions across edge and cloud computing centers.

\textbf{Edge interaction layer.}
The edge cloud layer provides low-latency data transmission and interaction services for the metaverse, supporting fast resource provisioning and service switching between different metaverse zones.
The edge cloud interaction layer contains three types of interaction, including ``edge-edge", ``edge-device", and ``edge-cloud" collaborations.
This layer manages multidimensional network resources in different metaverse zones, such as network, computing, storage, and other interaction demands. 
More importantly, users can shuttle between different metaverse zones and quickly switch between services through cloud-edge-device collaboration architecture, with reduced real delay in user interactions.

\textbf{Cloud interaction layer.}
The remote cloud center is generally far from the physical distance of the pervasive terminal but has ample computing resources and can provide the ability to connect different cloud centers and schedule edge cloud services.
For example, we can use the remote cloud to render complex scenes that are static or slow to update in the metaverse zone, and frequent dynamic rendering computations and user interaction commands can be scheduled on the edge and devices.
Besides, we can leverage the cloud centers to collaborate and schedule a larger range of SAGIN sensing data aggregation, computation, and digital twin reconstruction.

This article focuses on the basic multi-player interaction architecture for the metaverse and provides a basic platform for exploring more detailed scheduling solutions.
To this end, we implement a multi-player interaction prototype system in Figure~3.
Users can use XR glasses, laptops, smartphones, or terminals with cameras to access the prototype metaverse.
In our prototype, users can construct their avatar and drive it the prototype.
Besides, wearing XR glasses can acquire more channels to interact with others.
Next, we detail the multi-player interaction in our prototype, including the following two cases.
\paragraph{\textbf{Interaction in a single metaverse zone}}
A more general interaction scenario is to provide low-latency multi-player interaction services in the same metaverse zone.
As shown in Figure.~3, the multi-player interaction service requires stable connections, real-time response, and high-precision digital twin rendering for a single metaverse.
The interactions between various devices are mainly established around the edge cloud, including the D2D interaction and edge-assisted multi-player framework.
Taking the metaverse virtual meeting zone as an example, a basic static meeting scene is established and rendered in the edge cloud. 
Users can access it using different devices, such as smartphones, PCs, and XR glasses.
Then, the user can establish and synchronize a personal avatar according to the type of device, i.e., using XR glass can obtain more interaction experiences, such as gesture interaction.
Besides, when using D2D to synchronize interactive commands between users and initiate rendering update operations to the edge, it can quickly update and respond to the results to terminals. 
Thus, this interaction can greatly reduce the computing requirements of intelligent interactive terminals.
It is foreseeable that with the increase of access devices, the computing pressure and synchronization of the edge center will become more complicated. 
Therefore, we can enable part of the rendering and synchronization locally for the multi-player interaction in a single metaverse zone (i.e., the user with XR glass can receive interactive commands directly and locally from other devices and complete rendering)

\paragraph{\textbf{Interaction across multiple metaverse zones}}
Different users interact or switch identities across multiple metaverse zones may involve interactions between different terminals far away from the physical world, such as the interactions between different edge clouds in Figure.~3.
Thus, the metaverse should consider the stability of a single zone interaction and the interaction terminal with multiple avatar attributes, and how to switch and synchronize between different metaverse zones quickly.
In addition, the characteristics of different metaverse zone and interaction terminals are heterogeneous, requiring dynamic adaptation and optimal resource allocation to meet the interaction demands of each metaverse zone and each user.
Hence, designing a ``virtual interaction brain" between the edge cloud and the cloud center is necessary to optimize resource and task allocation dynamically.
For example, we can explore the scheduling method by integrating SDN/NFV, deep reinforcement learning, and multi-objective optimization technologies to implement efficient interactions.

\vspace{-0.2cm}
\subsection{Evaluation}
To evaluate the performance of the proposed cloud-edge-device collaborative system, we compared the rendering frame per second~(FPS) performance of synchronization and rendering between different multiplayer interaction methods, including 1) cloud-based interaction, 2) D2D-based interaction~\cite{ren2020edge}, and 3) edge-based interaction, respectively. 
We use several identical devices and access the 3D AR applications for online interaction, i.e., they connect to the same virtual meeting in Figure.~3 and interact with others separately.
We set all methods in the same WiFi network bandwidth and control the end-to-end latencies as 50ms, 10ms, and 15ms for the above three methods.

Figure.~4 and Figure.~5 evaluate our cloud-edge-device collaborative interaction with baseline methods.
Figure.~4 presents the real-time rendering FPS performance of different methods when increasing the number of access terminals.
We can see that increasing access users leads to a dramatic decrease in FPS of the D2D-based method due to the limited computing resource of the devices.
Also, the devices require maintaining many connections and data synchronization, resulting in the FPS decrease.
Hence, the D2D-based approach is stable access to services without going through the cloud and edge for a small number of interactions.
Besides, The cloud-based approach has a decrease in FPS with the increase of users' access, which is mainly due to the increasing cost of the synchronization and rendering computation.
Although the computing resources in the cloud center are sufficient, they still perform worse FPS compared to the edge-based approach due to a higher end-to-end interaction delay.
Last
The edge-based approach achieves decent FPS benefiting from the lower end-to-end delay. 
However, we know there is a significant increase in links and computation at the edge as the access users increase.
It is foreseeable that as access users continue to increase, the resources of edge computing will be continuously depleted, resulting in poor quality of service.
Therefore, the proposed cloud-edge-device collaboration approach can minimize computing resources and interaction latency.
\vspace{-0.2cm}
\begin{figure}[htbp]
	\centering
	\begin{minipage}[t]{0.24\textwidth}
		\centering
		\includegraphics[scale=0.2]{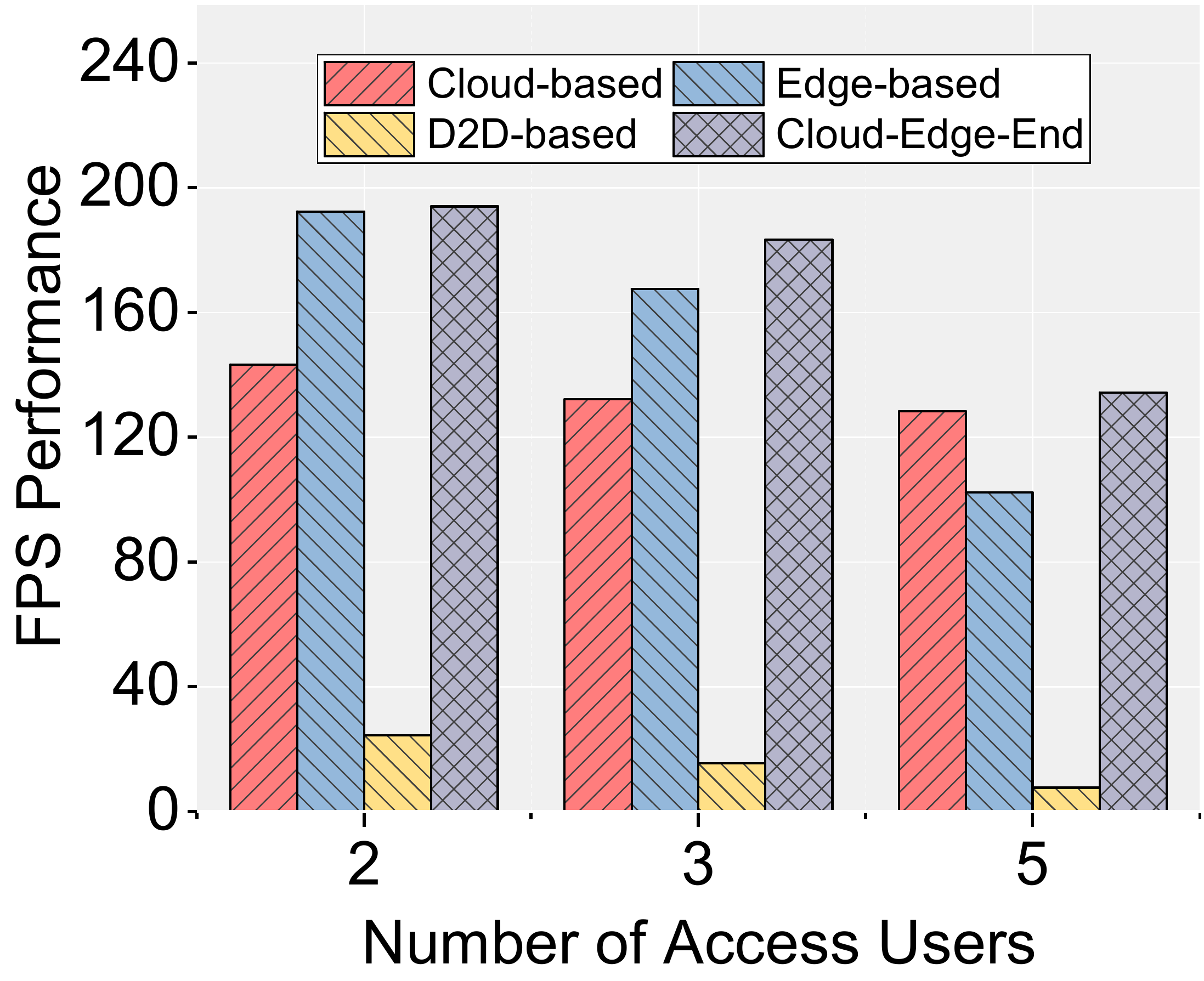}
		\caption{FPS comparison}
		\vspace{-0.2cm}
	\end{minipage}
	\begin{minipage}[t]{0.24\textwidth}
		\centering
		\includegraphics[scale=0.2]{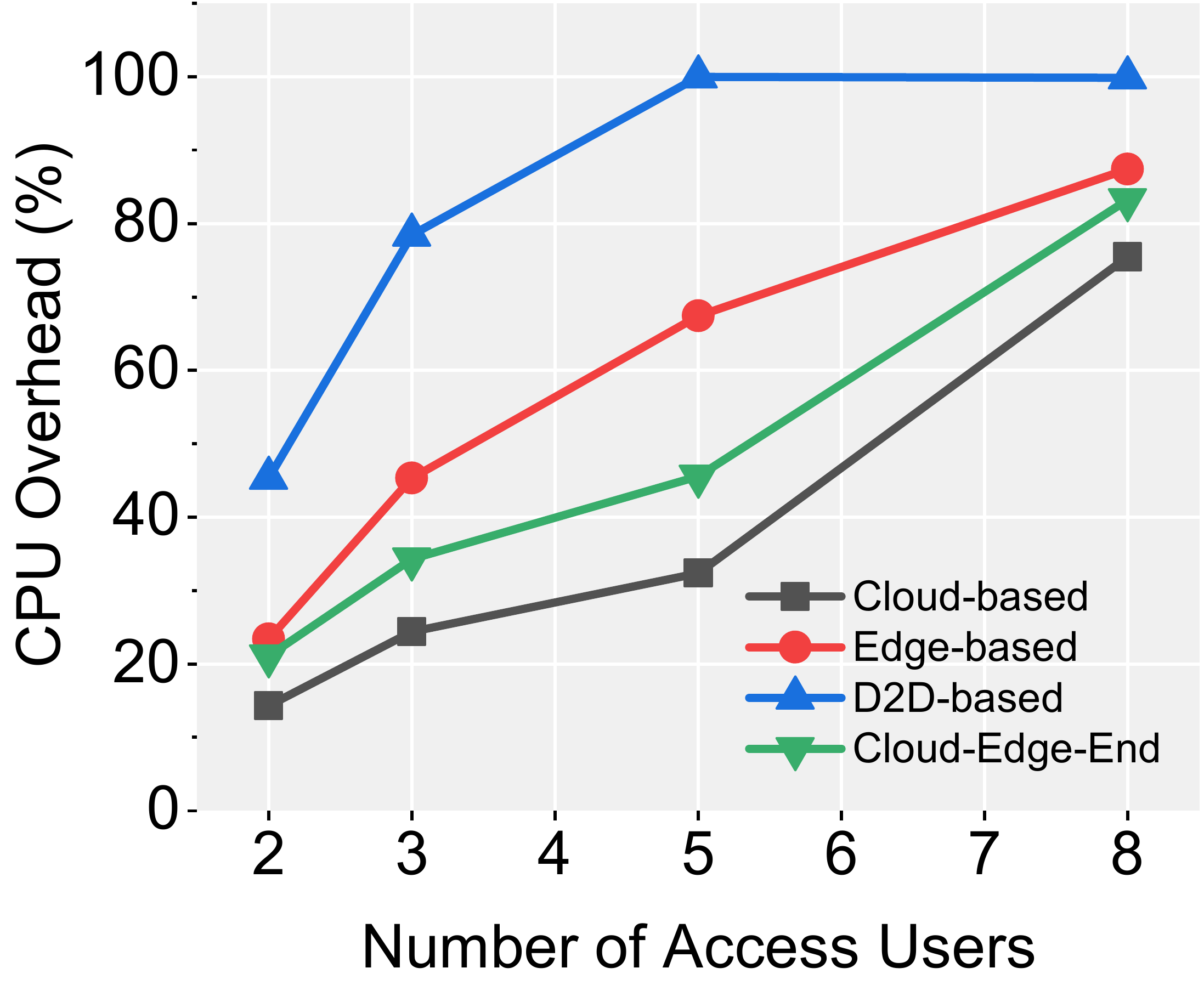}
		\caption{CPU performance}
				\vspace{-0.2cm}
	\end{minipage}
\end{figure}
Second, Figure.~5 presents the system CPU overhead of different interaction methods.
Similar to the FPS performance results, as access users increases, the range of use and the quality of multi-player interaction of the D2D approach decreases until it blocks the system's resources.
Then, cloud- and edge-based approaches lead to a dramatic increase in server resources in a short period of time.
When users leave the service or switch to other metaverse services, it again leads to an unsatisfactory waiting experience.

\vspace{-0.2cm}
\section{Future Research Directions}
We highlight further directions from the following aspects.
\paragraph{\textbf{Semantic communications for the metaverse}}
In the metaverse, transferring massive high-precision 3D models and 6-DoF videos (e.g., holographic videos) requires a Tbps lever of network bandwidth, far beyond the capacity of existing 5G/B5G.
Recently, semantic communication has shown a promising paradigm that reduces data transmission by transmitting key features and reconstructing the raw data with AI technologies~\cite{huang2022, huang2021aitransfer}.
Applying semantic communication technology to multi-player interactions can further improve efficiency and expand the scope of metaverse interactions.

\paragraph{\textbf{Redefine quality of experience~(QoE) of interactions}}
The metaverse contains a wide variety of heterogeneous services and the evaluation methods and metrics of these services are different.
Therefore, the QoE of metaverse interactions needs to be redefined, which requires an interdisciplinary and comprehensive definition from multiple perspectives. 
For example, we can evaluate objective and subjective from users' physiological perceptions and interaction experiences.
Last, it is better to correct differences between different services through a dynamic adaptive QoE evaluation framework that can self-learn and adjust.

\paragraph{\textbf{Incentives to stimulate users to participate}}
The metaverse expects users to actively participate and create rich digital assets, thus promoting the development of the metaverse. 
How to motivate different participants to actively participate in the metaverse system and promote it to other users, ultimately forming a well-developed ecosystem, is challenging.
This requires an in-depth analysis of the user behavior of the metaverse in the future and research on using methods that can enhance user participation. 

\paragraph{\textbf{Privacy and security}}
The metaverse will be built on a transactional infrastructure of blockchain technology, so as users continue to participate, a large amount of transactional information will be completed on the blockchain.
Thus, new security issues will emerge, including various cyber attacks.
For example, the increasing amount of user data (e.g., various physiological information) sensed by future metaverse interaction terminals may be collected for more accurate and personalized advertising, which will pose new challenges to user data privacy.

\paragraph{\textbf{Economics in metaverse}}
The metaverse opens up new possibilities for users and service providers to complete digital asset transactions in the virtual digital world.
The future of resources and the economy will extend from the physical to the digital world and may give rise to new economic models.
Thus exploring the metaverse and related economic implications is a new trend for the future.

\section{Conclusion}
Metaverse builds native technologys allowing us to interact and share realities together with our surroundings. In this article, we introduced the concept, characteristics, enabling techniques, and killer applications in the metaverse.
We then clarified the critical challenges of enabling multi-player interaction between heterogeneous terminals and various metaverse zones.
We present an implemented prototype metaverse multi-player interaction system based on a cloud-edge-device framework to provide flexibility, self-organization, and elastic interactions service.
We conducted experiments to illustrate the advantages of the proposed method compared with centralized- and D2D-based approaches.
Finally, we point out future research approaches and discuss potential solutions to improve immersive interaction in the metaverse.

\vspace{-0.2cm}
\section*{Acknowledgment}
This research was supported in part by the National Natural Science Foundation of China under Grant 62202065, in part by the Project funded by China Postdoctoral Science Foundation 2022TQ0047 and 2022M710465, and in part by Academy of Finland under projects 319670 and 326305.

\vspace{-0.2cm}
\bibliographystyle{IEEEtran}
\bibliography{IEEEabrv,IEEEexample}

\begin{thebibliography}{10}
\providecommand{\url}[1]{#1}
\csname url@samestyle\endcsname
\providecommand{\newblock}{\relax}
\providecommand{\bibinfo}[2]{#2}
\providecommand{\BIBentrySTDinterwordspacing}{\spaceskip=0pt\relax}
\providecommand{\BIBentryALTinterwordstretchfactor}{4}
\providecommand{\BIBentryALTinterwordspacing}{\spaceskip=\fontdimen2\font plus
\BIBentryALTinterwordstretchfactor\fontdimen3\font minus
  \fontdimen4\font\relax}
\providecommand{\BIBforeignlanguage}[2]{{%
\expandafter\ifx\csname l@#1\endcsname\relax
\typeout{** WARNING: IEEEtran.bst: No hyphenation pattern has been}%
\typeout{** loaded for the language `#1'. Using the pattern for}%
\typeout{** the default language instead.}%
\else
\language=\csname l@#1\endcsname
\fi
#2}}
\providecommand{\BIBdecl}{\relax}
\BIBdecl

\bibitem{tang2022roadmap}
F.~Tang, X.~Chen, M.~Zhao, and N.~Kato, ``The roadmap of communication and
  networking in 6g for the metaverse,'' \emph{IEEE Wireless Communications},
  2022.

\bibitem{wang2022survey}
Y.~Wang, Z.~Su, N.~Zhang, R.~Xing, D.~Liu, T.~H. Luan, and X.~Shen, ``A survey
  on metaverse: Fundamentals, security, and privacy,'' \emph{IEEE
  Communications Surveys \& Tutorials}, 2022.

\bibitem{zhao2022metaverse}
Y.~Zhao, J.~Jiang, Y.~Chen, R.~Liu, Y.~Yang, X.~Xue, and S.~Chen, ``Metaverse:
  Perspectives from graphics, interactions and visualization,'' \emph{Visual
  Informatics}, 2022.

\bibitem{yang2022fusing}
Q.~Yang, Y.~Zhao, H.~Huang, Z.~Xiong, J.~Kang, and Z.~Zheng, ``Fusing
  blockchain and ai with metaverse: A survey,'' \emph{IEEE Open Journal of the
  Computer Society}, vol.~3, pp. 122--136, 2022.

\bibitem{dionisio20133d}
J.~D.~N. Dionisio, W.~G.~B. III, and R.~Gilbert, ``3d virtual worlds and the
  metaverse: Current status and future possibilities,'' \emph{ACM Computing
  Surveys (CSUR)}, vol.~45, no.~3, pp. 1--38, 2013.

\bibitem{qiao2019web}
X.~Qiao, P.~Ren, S.~Dustdar, L.~Liu, H.~Ma, and J.~Chen, ``Web ar: A promising
  future for mobile augmented reality—state of the art, challenges, and
  insights,'' \emph{Proceedings of the IEEE}, vol. 107, no.~4, pp. 651--666,
  2019.

\bibitem{chen2021edge}
N.~Chen, T.~Qiu, L.~Zhao, X.~Zhou, and H.~Ning, ``Edge intelligent networking
  optimization for internet of things in smart city,'' \emph{IEEE Wireless
  Communications}, vol.~28, no.~2, pp. 26--31, 2021.

\bibitem{wang2021blockchain}
Y.~Wang, Z.~Su, J.~Ni, N.~Zhang, and X.~Shen, ``Blockchain-empowered
  space-air-ground integrated networks: Opportunities, challenges, and
  solutions,'' \emph{IEEE Communications Surveys \& Tutorials}, vol.~24, no.~1,
  pp. 160--209, 2021.

\bibitem{jankowski2013survey}
J.~Jankowski and M.~Hachet, ``A survey of interaction techniques for
  interactive 3d environments,'' in \emph{Eurographics 2013-STAR}, 2013.

\bibitem{bouzbib2021can}
E.~Bouzbib, G.~Bailly, S.~Haliyo, and P.~Frey, ``“can i touch this?”:
  Survey of virtual reality interactions via haptic solutions: Revue de
  litt{\'e}rature des interactions en r{\'e}alit{\'e} virtuelle par le biais de
  solutions haptiques,'' in \emph{Francophone sur l'Interaction Homme-Machine},
  2021, pp. 1--16.

\bibitem{ren2020edge}
P.~Ren, X.~Qiao, Y.~Huang, L.~Liu, C.~Pu, S.~Dustdar, and J.-L. Chen, ``Edge ar
  x5: An edge-assisted multi-user collaborative framework for mobile web
  augmented reality in 5g and beyond,'' \emph{IEEE Transactions on Cloud
  Computing}, 2020.

\bibitem{huang2022}
Y.~Huang, Y.~Zhu, X.~Qiao, X.~Su, S.~Dustdar, and P.~Zhang, ``Towards
  holographic video communications: A promising ai-driven solution,''
  \emph{IEEE IEEE Communication Magazine}, 2022.

\bibitem{huang2021aitransfer}
Y.~Huang, Y.~Zhu, X.~Qiao, Z.~Tan, and B.~Bai, ``Aitransfer: Progressive
  ai-powered transmission for real-time point cloud video streaming,'' in
  \emph{Proceedings of the 29th ACM International Conference on Multimedia},
  2021, pp. 3989--3997.

\end{thebibliography}

%
%
\vspace{-1cm}
\begin{IEEEbiographynophoto}
	{Yakun~Huang} is currently a Postdoctoral Researcher at the State Key Laboratory of Networking and Switching Technology, Beijing University of Posts and Telecommunications, Beijing, China.
	His current research interests include volumetric video streaming, mobile computing, and augmented reality.	
\end{IEEEbiographynophoto}
\vspace{-1cm}
\begin{IEEEbiographynophoto}
	{Xiuquan~Qiao } is currently a Full Professor with the State Key Laboratory of Networking and Switching Technology, Beijing University of Posts and Telecommunications, Beijing, China. His current research interests include the future Internet, services computing, computer vision, distributed deep learning, augmented reality, virtual reality, and 5G networks.
\end{IEEEbiographynophoto}
\vspace{-1cm}
\begin{IEEEbiographynophoto}
	{Haowen~Wang} is currently working towards a Ph.D. degree at the State Key Laboratory of Networking and Switching Technology, Beijing University of Posts and Telecommunications, Beijing, China. His current research interests include 3D object detection and deep learning.
\end{IEEEbiographynophoto}
\vspace{-1cm}
\begin{IEEEbiographynophoto}
	{Xiang~Su } is currently an Associate Professor with the Department of Computer Science, Norwegian University of Science and Technology, Norway, and the University of Oulu, Finland. He has extensive expertise in the Internet of Things, edge computing, mobile augmented reality, knowledge representations, and context modeling and reasoning. 
\end{IEEEbiographynophoto}
\vspace{-1cm}

\begin{IEEEbiographynophoto}
	{Schahram~Dustdar} (Fellow, IEEE) is a Full Professor of Computer Science and is heading the Distributed Systems Research Division at the TU Wien. He is an ACM Distinguished Scientist, ACM Distinguished Speaker, IEEE Fellow, and Member of Academia Europaea.
\end{IEEEbiographynophoto}
\vspace{-1cm}

\begin{IEEEbiographynophoto}
	{Ping~Zhang} (Fellow, IEEE) is currently a Full Professor and Director of the State Key Laboratory of Networking and Switching Technology, Beijing University of Posts and Telecommunications, Beijing, China.
	He is an Academician with the Chinese Academy of Engineering (CAE).
	He is also a member of the IMT-2020 (5G) Experts Panel and the Experts Panel for China’s 6G development.
\end{IEEEbiographynophoto}

\vfill

\end{document}